\documentclass[aps,preprint]{revtex4}
\usepackage{amssymb}
\usepackage{amsfonts}
\usepackage{amsmath}
\usepackage{graphicx}

\begin{document}

\title{Vulnerability and Defence: A Case for Stackelberg Game Dynamics}
\author{Azhar Iqbal$^{\dagger }$\thanks{%
Corresponding author, Email:azhar.iqbal@adelaide.edu.au, Phone: +61427661661.%
}, Ishan Honhaga$^{\dagger }$, Eyoel Teffera$^{\dagger \ddagger }$, Anthony
Perry$^{\ddagger }$, Robin Baker$^{\ddagger }$, Glenn Pearce$^{\ddagger }$,
and Claudia Szabo$^{\dagger }$}
\affiliation{$^{\dagger }$School of Computer and Mathematical Sciences, University of
Adelaide, South Australia 5005, Australia\\
$^{\ddagger }$Defence Science and Technology Group, Edinburgh, South
Australia, Australia.}

\begin{abstract}
This paper examines the tactical interaction between drones and tanks in
modern warfare through game theory, particularly focusing on Stackelberg
equilibrium and backward induction. It describes a high-stakes conflict
between two teams: one using advanced drones for attack, and the other
defending using tanks. The paper conceptualizes this as a sequential game,
illustrating the complex strategic dynamics similar to Stackelberg
competition, where moves and countermoves are carefully analyzed and
predicted.
\end{abstract}

\maketitle

\section{Introduction}

More than a century before John Nash formalized the concept of equilibrium in game theory \cite{Binmore, Rasmusen, Osborne}, Antoine Cournot \cite%
{Cournot1897} had already introduced a similar idea through his duopoly model, which became a cornerstone in the study of industrial organization \cite%
{Tirole1988}. In economics, an oligopoly refers to a market structure in which a small number of firms ($n\geq 2$) supply a particular product. A duopoly, a specific case where $n=2$, is the scenario to which Cournot's model applies. In this model, two firms simultaneously produce and sell a homogeneous product. Cournot identified an equilibrium quantity for each firm, where the optimal strategy for each participant is to follow a specific rule if the other firm adheres to it. This idea of equilibrium in a duopoly anticipated Nash’s more general concept of equilibrium points in non-cooperative games.

In 1934, Heinrich von Stackelberg \cite{Stackelberg1934,Gibbons1992} introduced a dynamic extension to Cournot’s model by allowing for sequential moves rather than simultaneous ones. In the Stackelberg model, one firm, the leader, moves first, while the second, the follower, reacts accordingly. A well-known example of such strategic behavior is General Motors’ leadership in the early U.S. automobile industry, with Ford and Chrysler often acting as followers.

The Stackelberg equilibrium, derived through backward induction, represents the optimal outcome in these sequential-move games. This equilibrium is often considered more robust than Nash equilibrium (NE) in such settings, as sequential games can feature multiple NEs, but only one corresponds to the backward-induction outcome \cite{Binmore, Rasmusen, Osborne}.

\section{Related Work}

Stackelberg games have been significantly influential in security and
military research applications \cite%
{Bustamante2024,Hunt2024,Chen2022,Bansal2021,Li2019,Feng2019,Sinha2018,Kar2016,Tambe2011,Paruchuri2008,Korzhyk2011,Hohzaki2009}%
. These games, based on the Stackelberg competition model, have been
successfully applied in a wide range of real-world scenarios. They are particularly
notable for their deployment in contexts where security decisions are
critical, such as in protecting infrastructure and managing military
operations.

The sequential setup of Stackelberg games is particularly
relevant in military contexts where strategic decisions often involve
anticipating and responding to an adversary's actions. The applications of
such games in military settings are diverse, ranging from optimizing
resource allocation for defence to strategizing offensive maneuvers.

This paper considers the strategic interplay between drones and tanks
through the lens of the Stackelberg equilibrium and the principles of
backward induction. In a military operation setting, we consider two types
of agents, namely, the attacker (Red team) and the defender (Blue team). The attacker might utilize mobile threats to attack or
reduce the number of static, unmovable entities belonging to the defender. 
% % conflict zones, we consider two opposing factions are
% locked in a high-stakes tactical duel. Drones, equipped with advanced
% surveillance and precision strike abilities, are utilized by one side. 
In response, the defender will employ some countermeasures to reduce the
number of enemy attackers. 
% employs heavily armored tanks known for their formidable defence and firepower. 
This complex pattern of strategic moves and countermoves is explored as a
sequential game, drawing on the concept of Stackelberg competition to
illuminate the dynamics at play.

While focusing on developing a game-theoretical analysis, we have presented a hypothetical strategic scenario involving tanks and drones to illustrate our point. Naturally, this scenario may only loosely reflect the realities of such encounters, which evolve rapidly and are subject to constant change.

This paper's contribution consists of obtaining an analytical solution to a
Stackelberg competition in a military setting. To obtain such a solution, we
limit the number of available strategic moves to small numbers but enough
still to demonstrate the dynamics of a sequential strategic military
operation.

\section{The game definition}

We consider a scenario in which two teams, Blue ($\mathcal{B}$) and Red ($%
\mathcal{R}$), are engaged in a military operation. The $\mathcal{B}$ team
comprises ground units, specifically tanks, while the $\mathcal{R}$ team
operates aerial units, namely drones. The strategy involves the $\mathcal{R}$
team's drones targeting the $\mathcal{B}$ team's tanks. Meanwhile, the $%
\mathcal{B}$ team not only has the capability to shoot down these drones but
also provides defensive cover for their tanks, creating a complex interplay
of offensive and defensive maneuvers in this combat scenario.

We assume that the $\mathcal{B}$ team consists of $n$ tanks where $n\in 
\mathcal{N}$ and represent the set of tanks as $T=\{\mathfrak{T}_{1},%
\mathfrak{T}_{2},...\mathfrak{T}_{n}\}.$ Let $S=\{\mathcal{S}_{1},\mathcal{S}%
_{2},...\mathcal{S}_{m}\}$ be a set of resources that are at the disposal of 
$\mathcal{B}$ team to protect the tanks. It is assumed that the $\mathcal{R}$
team's pure strategy is to attack one of the tanks from the set $T$. The $%
\mathcal{R}$ team's mixed strategy is then defined as a vector $\left\langle 
{\normalsize A}_{\mathfrak{T}}\right\rangle $ where ${\normalsize A}_{%
\mathfrak{T}}$ is the probability of attacking the tank $\mathfrak{T}:%
{\normalsize A}\rightarrow $ attack and $\sum\limits_{\mathfrak{T}=1}^{_{n}}%
{\normalsize A}_{\mathfrak{T}}=1.$ The $\mathcal{B}$ team's mixed strategy
is also a vector $\left\langle D_{\mathfrak{T}}\right\rangle $ where $D_{%
\mathfrak{T}}$ is the marginal probability of protecting the tank $\mathfrak{%
T}$. Note that a marginal is a sum of probabilities and it contrasts with a
conditional distribution which gives the probabilities contingent upon the
values of the other variables.

\subsection{Marginals of supporting resources for the tanks}

We consider the case when there are five resources from the set $\{\mathcal{S%
}_{1},\mathcal{S}_{2},...\mathcal{S}_{5}\}$ available to protect the four
tanks from $T=\{\mathfrak{T}_{1},\mathfrak{T}_{2},...\mathfrak{T}_{4}\},$
whereas one or more of the resources can be used to protect a tank $%
\mathfrak{T}_{i}$ from the set $T$. For this case, the marginal
probabilities $D_{\mathfrak{T}}$ to protect the tank $\mathfrak{T}$ are
determined as follows:

\begin{equation}
\begin{array}{c}
\text{Tanks}%
\end{array}%
\overset{%
\begin{array}{c}
\text{Resources}%
\end{array}%
}{%
\begin{array}{ccccccc}
& {\small \mathcal{S}}_{1} & {\small \mathcal{S}}_{2} & {\small \mathcal{S}}%
_{3} & {\small \mathcal{S}}_{4} & {\small \mathcal{S}}_{5} & D_{\mathfrak{T}}
\\ 
{\small \mathfrak{T}}_{1} & \Pr {\small (\mathfrak{T}}_{1}{\small ,\mathcal{S%
}}_{1}{\small )} & \Pr {\small (\mathfrak{T}}_{1}{\small ,\mathcal{S}}_{2}%
{\small )} & \Pr {\small (\mathfrak{T}}_{1}{\small ,\mathcal{S}}_{3}{\small )%
} & \Pr {\small (\mathfrak{T}}_{1}{\small ,\mathcal{S}}_{4}{\small )} & \Pr 
{\small (\mathfrak{T}}_{1}{\small ,\mathcal{S}}_{5}{\small )} & {\small D}%
_{1}{\small =}\frac{1}{5}\sum\limits_{i=1}^{5}\Pr {\small (\mathfrak{T}}_{1}%
{\small ,\mathcal{S}}_{i}{\small )} \\ 
{\small \mathfrak{T}}_{2} & \Pr {\small (\mathfrak{T}}_{2}{\small ,\mathcal{S%
}}_{1}{\small )} & \Pr {\small (\mathfrak{T}}_{2}{\small ,\mathcal{S}}_{2}%
{\small )} & \Pr {\small (\mathfrak{T}}_{2}{\small ,\mathcal{S}}_{3}{\small )%
} & \Pr {\small (\mathfrak{T}}_{2}{\small ,\mathcal{S}}_{4}{\small )} & \Pr 
{\small (\mathfrak{T}}_{2}{\small ,\mathcal{S}}_{5}{\small )} & {\small D}%
_{2}{\small =}\frac{1}{5}\sum\limits_{i=1}^{5}\Pr {\small (\mathfrak{T}}_{2}%
{\small ,\mathcal{S}}_{i}{\small )} \\ 
{\small \mathfrak{T}}_{3} & \Pr {\small (\mathfrak{T}}_{3}{\small ,\mathcal{S%
}}_{1}{\small )} & \Pr {\small (\mathfrak{T}}_{3}{\small ,\mathcal{S}}_{2}%
{\small )} & \Pr {\small (\mathfrak{T}}_{3}{\small ,\mathcal{S}}_{3}{\small )%
} & \Pr {\small (\mathfrak{T}}_{3}{\small ,\mathcal{S}}_{4}{\small )} & \Pr 
{\small (\mathfrak{T}}_{3}{\small ,\mathcal{S}}_{5}{\small )} & {\small D}%
_{3}{\small =}\frac{1}{5}\sum\limits_{i=1}^{5}\Pr {\small (\mathfrak{T}}_{3}%
{\small ,\mathcal{S}}_{i}{\small )} \\ 
{\small \mathfrak{T}}_{4} & \Pr {\small (\mathfrak{T}}_{4}{\small ,\mathcal{S%
}}_{1}{\small )} & \Pr {\small (\mathfrak{T}}_{4}{\small ,\mathcal{S}}_{2}%
{\small )} & \Pr {\small (\mathfrak{T}}_{4}{\small ,\mathcal{S}}_{3}{\small )%
} & \Pr {\small (\mathfrak{T}}_{4}{\small ,\mathcal{S}}_{4}{\small )} & \Pr 
{\small (\mathfrak{T}}_{4}{\small ,\mathcal{S}}_{5}{\small )} & {\small D}%
_{4}{\small =}\frac{1}{5}\sum\limits_{i=1}^{5}\Pr {\small (\mathfrak{T}}_{4}%
{\small ,\mathcal{S}}_{i}{\small )}%
\end{array}%
,}
\end{equation}%
where

\begin{gather}
\sum\limits_{i=1}^{4}\Pr (\mathfrak{T}_{i},\mathcal{S}_{1})=\sum%
\limits_{i=1}^{4}\Pr (\mathfrak{T}_{i},\mathcal{S}_{2})=\sum%
\limits_{i=1}^{4}\Pr (\mathfrak{T}_{i},\mathcal{S}_{3})  \notag \\
=\sum\limits_{i=1}^{4}\Pr (\mathfrak{T}_{i},\mathcal{S}_{4})=\sum%
\limits_{i=1}^{4}\Pr (\mathfrak{T}_{i},\mathcal{S}_{5})=1.
\end{gather}%
For instance, $\Pr (\mathfrak{T}_{3},\mathcal{S}_{2})$ is the probability
that the resource $\mathcal{S}_{2}$ is used to give protection to the tank $%
\mathfrak{T}_{3}.$ This means $D_{\mathfrak{i}}\leq 1$ and $\sum\limits_{%
\mathfrak{i}=1}^{4}D_{\mathfrak{i}}=1.$ In case the set of resources
consists of $m$ element i.e. $S=\{\mathcal{S}_{1},\mathcal{S}_{2},...%
\mathcal{S}_{m}\}$ and the number of tanks to be protected are $n$, we then
have the the marginal probabilities $D_{\mathfrak{j}}$ to protect the $j$-th
tank obtained as ${\small D}_{j}{\small =}\frac{1}{m}\sum\limits_{i=1}^{m}%
\Pr {\small (\mathfrak{T}}_{j}{\small ,\mathcal{S}}_{i}{\small )}$ where $%
1\leq j\leq n.$

\subsection{Defining the Reward Functions}

Let $\mathfrak{R}_{\mathcal{B}}(\mathfrak{T})$ be the reward to team $%
\mathcal{B}$ if attacked tank $\mathfrak{T}$ is protected using resources
from the set $S=\{\mathcal{S}_{1},\mathcal{S}_{2},...\mathcal{S}_{m}\},$ $%
\mathfrak{C}_{\mathcal{B}}(\mathfrak{T})$ be the cost to team $\mathcal{B}$
if attacked tank $\mathfrak{T}$ is unprotected, $\mathfrak{R}_{\mathcal{R}}(%
\mathfrak{T})$ be the reward to team $\mathcal{R}$ if attacked tank $%
\mathfrak{T}$ is unprotected, $\mathfrak{C}_{\mathcal{R}}(\mathfrak{T})$ be
the cost to team $\mathcal{R}$ if attacked tank $\mathfrak{T}$ is protected.
Note that $D_{\mathfrak{T}}$ is the marginal of protecting the tank $%
\mathfrak{T}$ using the resources from the set $S.$

The quantity $D_{\mathfrak{T}}\mathfrak{R}_{\mathcal{B}}(\mathfrak{T})-(1-D_{%
\mathfrak{T}})\mathfrak{C}_{\mathcal{B}}(\mathfrak{T})$ then describes the
payoff to the $\mathcal{B}$ team when tank $\mathfrak{T}$ is attacked.
Similarly, the quantity $(1-D_{\mathfrak{T}})\mathfrak{R}_{\mathcal{R}}(%
\mathfrak{T})-D_{\mathfrak{T}}\mathfrak{C}_{\mathcal{R}}(\mathfrak{T})$
describes the payoff to the $\mathcal{R}$ team when the tank $\mathfrak{T}$
is attacked. However, the probability that the tank $\mathfrak{T}$ is
attacked is ${\normalsize A}_{\mathfrak{T}}$ and we can take this into
consideration to define the quantities ${\normalsize A}_{\mathfrak{T}}\{D_{%
\mathfrak{T}}\mathfrak{R}_{\mathcal{B}}(\mathfrak{T})-(1-D_{\mathfrak{T}})%
\mathfrak{C}_{\mathcal{B}}(\mathfrak{T})\}$ and ${\normalsize A}_{\mathfrak{T%
}}\{(1-D_{\mathfrak{T}})\mathfrak{R}_{\mathcal{R}}(\mathfrak{T})-D_{%
\mathfrak{T}}\mathfrak{C}_{\mathcal{R}}(\mathfrak{T})\}$. These are the
contributions to the payoffs to the $\mathcal{B}$ and $\mathfrak{T}$ teams,
respectively, when the tank $\mathfrak{T}$ is attacked with the probability $%
{\normalsize A}_{\mathfrak{T}}.$

As the vector $\left\langle {\normalsize A}_{\mathfrak{T}}\right\rangle $
describes the $\mathcal{R}$ team's (mixed) attacking strategy whereas the
vector $\left\langle D_{\mathfrak{T}}\right\rangle $ describes the $\mathcal{%
B}$ team's (mixed) protection strategy, the players' strategy profiles are
given as $\{\left\langle D_{\mathfrak{T}}\right\rangle,\left\langle 
{\normalsize A}_{\mathfrak{T}}\right\rangle \}$. For a set of tanks $T$, the
expected payoffs \cite{Tambe2011, Paruchuri2008} to the $\mathcal{B}$ and $\mathcal{R}$ teams, respectively,
can then be written as

\begin{eqnarray}
\Pi _{\mathcal{B}}\{\left\langle D_{\mathfrak{T}}\right\rangle ,\left\langle 
{\normalsize A}_{\mathfrak{T}}\right\rangle \} &=&\sum\limits_{\mathfrak{T}%
\in T}{\normalsize A}_{\mathfrak{T}}\{D_{\mathfrak{T}}\mathfrak{R}_{\mathcal{%
B}}(\mathfrak{T})-(1-D_{\mathfrak{T}})\mathfrak{C}_{\mathcal{B}}(\mathfrak{T}%
)\},  \notag \\
\Pi _{\mathcal{R}}\{\left\langle D_{\mathfrak{T}}\right\rangle ,\left\langle 
{\normalsize A}_{\mathfrak{T}}\right\rangle \} &=&\sum\limits_{\mathfrak{T}%
\in T}{\normalsize A}_{\mathfrak{T}}\{(1-D_{\mathfrak{T}})\mathfrak{R}_{%
\mathcal{R}}(\mathfrak{T})-D_{\mathfrak{T}}\mathfrak{C}_{\mathcal{R}}(%
\mathfrak{T})\}.  \label{Reward_functions}
\end{eqnarray}

We note from these payoffs that if the attack probability for a tank $%
\mathfrak{T}$ is zero, the rewards to both the $\mathcal{B}$ and $\mathfrak{T%
}$ teams for the tank $\mathfrak{T\in }$\ $T$ are also zero; the payoff
functions for the either team depend only on the attacked tanks; and if the $%
\mathcal{B}$ and $\mathcal{R}$ teams move simultaneously, the solution is a
Nash equilibrium.

Note that with reference to the reward functions defined in Eqs. (\ref%
{Reward_functions}), for the imagined strategy profile (not equilibrium),
where defender protects the first tank by all elements from the set $\{%
\mathcal{S}_{1},\mathcal{S}_{2},...\mathcal{S}_{5}\}$ i.e. $\Pr {\small (%
\mathfrak{T}}_{1}{\small ,\mathcal{S}}_{i}{\small )=1}$ for $1\leq i\leq 5,$
we obtain ${\small D}_{1}{\small =}\frac{1}{5}\sum\limits_{i=1}^{5}\Pr 
{\small (\mathfrak{T}}_{1}{\small ,\mathcal{S}}_{i}{\small )=1}$ and thus $%
{\small D}_{2,3,4}=0.$ If the attacker decides to attack the first tank i.e. 
${\normalsize A}_{\mathfrak{1}}=1$, we obtain $\Pi _{\mathcal{B}%
}\{\left\langle D_{\mathfrak{T}}\right\rangle ,\left\langle {\normalsize A}_{%
\mathfrak{T}}\right\rangle \}=\mathfrak{R}_{\mathcal{B}}(\mathfrak{T}_{1})$
and $\Pi _{\mathcal{R}}\{\left\langle D_{\mathfrak{T}}\right\rangle
,\left\langle {\normalsize A}_{\mathfrak{T}}\right\rangle \}=-\mathfrak{C}_{%
\mathcal{R}}(\mathfrak{T}_{1}).$

\section{Leader-Follower Interaction and Stackelberg Equilibrium}

We consider a three step strategic game between the $\mathcal{B}$ and $%
\mathcal{R}$ teams, also called the leader-follower interaction. As the
leader, the $\mathcal{B}$ team chooses an action consisting of a protection
strategy $\left\langle D_{\mathfrak{T}}\right\rangle $. The $\mathcal{R}$
team observes $\left\langle D_{\mathfrak{T}}\right\rangle $ and then chooses
an action consisting of its attack strategy given by the vector $%
\left\langle {\normalsize A}_{\mathfrak{T}}\right\rangle $. Knowing the
rational response $\left\langle {\normalsize A}_{\mathfrak{T}}\right\rangle $
of the $\mathcal{R}$ team, the $\mathcal{B}$ team takes this in account and
as the leader optimizes its own action. The payoffs to the two teams are $%
\Pi _{\mathcal{B}}\{\left\langle D_{\mathfrak{T}}\right\rangle ,\left\langle 
{\normalsize A}_{\mathfrak{T}}\right\rangle \}$ and $\Pi _{\mathcal{R}%
}\{\left\langle D_{\mathfrak{T}}\right\rangle ,\left\langle {\normalsize A}_{%
\mathfrak{T}}\right\rangle \}$.

This game is an example of the dynamic games
of complete and perfect information \cite{Rasmusen}. Key features of this
game are a) the moves occur in sequence, b) all previous moves are known
before next move is chosen, and c) the players' payoffs are common
knowledge. This framework allows for strategic decision-making based on the
actions and expected reactions of the other players, typical of Stackelberg
competition scenarios. In many real-world scenarios---especially in complex
environments in a military contexts---the assumption that players' payoffs
are common knowledge does not hold and complete information about the
payoffs of other players is rarely available.

Given the action $\left\langle D_{\mathfrak{T}}\right\rangle $ is previously
chosen by the $\mathcal{B}$ team, at the second stage of the game, when the $%
\mathcal{R}$ team gets the move, it faces the problem:

\begin{equation}
\underset{\left\langle {\normalsize A}_{\mathfrak{T}}\right\rangle }{Max}%
\text{ }\Pi _{\mathcal{R}}\{\left\langle D_{\mathfrak{T}}\right\rangle
,\left\langle {\normalsize A}_{\mathfrak{T}}\right\rangle \}  \label{max2}
\end{equation}%
Assume that for each $\left\langle D_{\mathfrak{T}}\right\rangle $, $%
\mathcal{R}$ team's optimization problem (\ref{max2}) has a unique solution $%
\mathcal{S}_{\mathcal{R}}(\left\langle D_{\mathfrak{T}}\right\rangle )$,
which is known as the \textit{best response} of the $\mathcal{R}$ team. Now
the $\mathcal{B}$ team can also solve the $\mathcal{R}$ team's optimization
problem by anticipating the $\mathcal{R}$ team's response to each action $%
\left\langle D_{\mathfrak{T}}\right\rangle $ that the $\mathcal{B}$ team
might take. So that the $\mathcal{B}$ team faces the problem:

\begin{equation}
\underset{\left\langle D_{\mathfrak{T}}\right\rangle }{Max}\text{ }\Pi _{%
\mathcal{B}}\{\left\langle D_{\mathfrak{T}}\right\rangle ,\mathcal{S}_{%
\mathcal{R}}(\left\langle D_{\mathfrak{T}}\right\rangle )\}  \label{max1}
\end{equation}%
Suppose this optimization problem also has a unique solution for the $%
\mathcal{B}$ team and is denoted by $\left\langle D_{\mathfrak{T}%
}\right\rangle ^{\ast }$. The solution $(\left\langle D_{\mathfrak{T}%
}\right\rangle ^{\ast },\mathcal{S}_{\mathcal{R}}(\left\langle D_{\mathfrak{T%
}}\right\rangle ^{\ast }))$ is the \textit{backwards-induction outcome} of
this game.

To address this, we consider the above simplified case i.e. when $\mathfrak{T%
}=1,2,..4$. Expanding Eq. (\ref{Reward_functions}) we obtain:%
\begin{equation}
\Pi _{\mathcal{R}}\{\left\langle D_{\mathfrak{T}}\right\rangle ,\left\langle 
{\normalsize A}_{\mathfrak{T}}\right\rangle \}=\sum\limits_{i=1}^{4}%
{\normalsize A}_{i}\{(1-D_{i})\mathfrak{R}_{\mathcal{R}}(\mathfrak{T}%
_{i})-D_{i}\mathfrak{C}_{\mathcal{R}}(\mathfrak{T}_{i})\}.
\label{Red_payoffs_expanded}
\end{equation}

Now, as $\sum\limits_{\mathfrak{T}=1}^{4}{\normalsize A}_{\mathfrak{T}}=1,$,
we take as an arbitrary choice ${\normalsize A}_{3}=1-{\normalsize A}_{1}-%
{\normalsize A}_{2}-{\normalsize A}_{4}$ in Eq. (\ref{Red_payoffs_expanded})
to obtain

\begin{gather}
\Pi _{\mathcal{R}}\{\left\langle D_{\mathfrak{T}}\right\rangle ,\left\langle 
{\normalsize A}_{\mathfrak{T}}\right\rangle \}=\sum\limits_{\substack{ i=1
\\ i\neq 3}}^{4}{\normalsize A}_{i}\{(1-D_{i})\mathfrak{R}_{\mathcal{R}}(%
\mathfrak{T}_{i})-D_{i}\mathfrak{C}_{\mathcal{R}}(\mathfrak{T}_{i})\}  \notag
\\
+(1-{\normalsize A}_{1}-{\normalsize A}_{2}-{\normalsize A}_{4})\{(1-D_{3})%
\mathfrak{R}_{\mathcal{R}}(\mathfrak{T}_{3})-D_{3}\mathfrak{C}_{\mathcal{R}}(%
\mathfrak{T}_{3})\},
\end{gather}%
and this re-presses the $\mathcal{R}$ team's reward function in terms of
only three variables ${\normalsize A}_{1},$ ${\normalsize A}_{2},$ and $%
{\normalsize A}_{3}$---defining its attack strategy $\left\langle 
{\normalsize A}_{\mathfrak{T}}\right\rangle .$ When expanded, the above
equation becomes

\begin{gather}
\Pi _{\mathcal{R}}\{\left\langle D_{\mathfrak{T}}\right\rangle ,\left\langle 
{\normalsize A}_{\mathfrak{T}}\right\rangle \}={\normalsize A}_{1}\{(1-D_{1})%
\mathfrak{R}_{\mathcal{R}}(\mathfrak{T}_{1})-D_{1}\mathfrak{C}_{\mathcal{R}}(%
\mathfrak{T}_{1})-(1-D_{3})\mathfrak{R}_{\mathcal{R}}(\mathfrak{T}_{3})+D_{3}%
\mathfrak{C}_{\mathcal{R}}(\mathfrak{T}_{3})\}  \notag \\
+{\normalsize A}_{2}\{(1-D_{2})\mathfrak{R}_{\mathcal{R}}(\mathfrak{T}%
_{2})-D_{2}\mathfrak{C}_{\mathcal{R}}(\mathfrak{T}_{2})-(1-D_{3})\mathfrak{R}%
_{\mathcal{R}}(\mathfrak{T}_{3})+D_{3}\mathfrak{C}_{\mathcal{R}}(\mathfrak{T}%
_{3})\}  \notag \\
+{\normalsize A}_{4}\{(1-D_{4})\mathfrak{R}_{\mathcal{R}}(\mathfrak{T}%
_{4})-D_{4}\mathfrak{C}_{\mathcal{R}}(\mathfrak{T}_{4})-(1-D_{3})\mathfrak{R}%
_{\mathcal{R}}(\mathfrak{T}_{3})+D_{3}\mathfrak{C}_{\mathcal{R}}(\mathfrak{T}%
_{3})\}  \notag \\
+(1-D_{3})\mathfrak{R}_{\mathcal{R}}(\mathfrak{T}_{3})-D_{3}\mathfrak{C}_{%
\mathcal{R}}(\mathfrak{T}_{3}).  \label{R_team_payoff}
\end{gather}%
Being a rational player, the $\mathcal{B}$ team knows that the $\mathcal{R}$
team would maximize its reward function with respect to its strategic
variables and this is expressed as%
\begin{equation}
\frac{\partial \Pi _{\mathcal{R}}\{\left\langle D_{\mathfrak{T}%
}\right\rangle ,\left\langle {\normalsize A}_{\mathfrak{T}}\right\rangle \}}{%
\partial {\normalsize A}_{1}}=\frac{\partial \Pi _{\mathcal{R}%
}\{\left\langle D_{\mathfrak{T}}\right\rangle ,\left\langle {\normalsize A}_{%
\mathfrak{T}}\right\rangle \}}{\partial {\normalsize A}_{2}}=\frac{\partial
\Pi _{\mathcal{R}}\{\left\langle D_{\mathfrak{T}}\right\rangle ,\left\langle 
{\normalsize A}_{\mathfrak{T}}\right\rangle \}}{\partial {\normalsize A}_{4}}%
=0,  \label{Optimization}
\end{equation}%
where ${\normalsize A}_{1},{\normalsize A}_{2},{\normalsize A}_{4}\in
\lbrack 0,1]$ and $\sum\limits_{\mathfrak{T}=1}^{4}{\normalsize A}_{%
\mathfrak{T}}=1.$ This results in obtaining%
\begin{gather}
D_{3}\left\{ \mathfrak{R}_{\mathcal{R}}(\mathfrak{T}_{3})+\mathfrak{C}_{%
\mathcal{R}}(\mathfrak{T}_{3})\right\} =D_{1}\left\{ \mathfrak{R}_{\mathcal{R%
}}(\mathfrak{T}_{1})+\mathfrak{C}_{\mathcal{R}}(\mathfrak{T}_{1})\right\} -%
\mathfrak{R}_{\mathcal{R}}(\mathfrak{T}_{1})+\mathfrak{R}_{\mathcal{R}}(%
\mathfrak{T}_{3}),  \notag \\
D_{3}\left\{ \mathfrak{R}_{\mathcal{R}}(\mathfrak{T}_{3})+\mathfrak{C}_{%
\mathcal{R}}(\mathfrak{T}_{3})\right\} =D_{2}\left\{ \mathfrak{R}_{\mathcal{R%
}}(\mathfrak{T}_{2})+\mathfrak{C}_{\mathcal{R}}(\mathfrak{T}_{2})\right\} -%
\mathfrak{R}_{\mathcal{R}}(\mathfrak{T}_{2})+\mathfrak{R}_{\mathcal{R}}(%
\mathfrak{T}_{3}),  \notag \\
D_{3}\left\{ \mathfrak{R}_{\mathcal{R}}(\mathfrak{T}_{3})+\mathfrak{C}_{%
\mathcal{R}}(\mathfrak{T}_{3})\right\} =D_{4}\left\{ \mathfrak{R}_{\mathcal{R%
}}(\mathfrak{T}_{4})+\mathfrak{C}_{\mathcal{R}}(\mathfrak{T}_{4})\right\} -%
\mathfrak{R}_{\mathcal{R}}(\mathfrak{T}_{4})+\mathfrak{R}_{\mathcal{R}}(%
\mathfrak{T}_{3}),  \label{conditions}
\end{gather}%
and this leads us to denote the sum of the reward and the cost to the $%
\mathcal{B}$ and $\mathcal{R}$ teams for protecting or attacking the tank $%
\mathfrak{T}$, respectively, by new symbols

\begin{eqnarray}
\Omega _{1}^{\mathcal{R}} &=&\mathfrak{R}_{\mathcal{R}}(\mathfrak{T}_{1})+%
\mathfrak{C}_{\mathcal{R}}(\mathfrak{T}_{1}),\text{ }\Omega _{2}^{\mathcal{R}%
}=\mathfrak{R}_{\mathcal{R}}(\mathfrak{T}_{2})+\mathfrak{C}_{\mathcal{R}}(%
\mathfrak{T}_{2}),  \notag \\
\Omega _{3}^{\mathcal{R}} &=&\mathfrak{R}_{\mathcal{R}}(\mathfrak{T}_{3})+%
\mathfrak{C}_{\mathcal{R}}(\mathfrak{T}_{3}),\text{ }\Omega _{4}^{\mathcal{R}%
}=\mathfrak{R}_{\mathcal{R}}(\mathfrak{T}_{4})+\mathfrak{C}_{\mathcal{R}}(%
\mathfrak{T}_{4}),  \notag \\
\Omega _{1}^{\mathcal{B}} &=&\mathfrak{R}_{\mathcal{B}}(\mathfrak{T}_{1})+%
\mathfrak{C}_{\mathcal{B}}(\mathfrak{T}_{1}),\text{ }\Omega _{2}^{\mathcal{B}%
}=\mathfrak{R}_{\mathcal{B}}(\mathfrak{T}_{2})+\mathfrak{C}_{\mathcal{B}}(%
\mathfrak{T}_{2}),  \notag \\
\Omega _{3}^{\mathcal{B}} &=&\mathfrak{R}_{\mathcal{B}}(\mathfrak{T}_{3})+%
\mathfrak{C}_{\mathcal{B}}(\mathfrak{T}_{3}),\text{ }\Omega _{4}^{\mathcal{B}%
}=\mathfrak{R}_{\mathcal{B}}(\mathfrak{T}_{4})+\mathfrak{C}_{\mathcal{B}}(%
\mathfrak{T}_{4}).  \label{substitutions}
\end{eqnarray}%
As $D_{\mathfrak{i}}\leq 5$ and $\sum\limits_{\mathfrak{i}=1}^{4}D_{%
\mathfrak{i}}=1$, we substitute $D_{3}=(1-D_{1}-D_{2}-D_{4})$ in Eqs. (\ref%
{conditions}) along with the substitutions (\ref{substitutions}) to obtain

\begin{gather}
(1-D_{2}-D_{4})\Omega _{3}^{\mathcal{R}}=D_{1}\left\{ \Omega _{1}^{\mathcal{R%
}}+\Omega _{3}^{\mathcal{R}}\right\} -\mathfrak{R}_{\mathcal{R}}(\mathfrak{T}%
_{1})+\mathfrak{R}_{\mathcal{R}}(\mathfrak{T}_{3}),  \label{1_} \\
(1-D_{1}-D_{4})\Omega _{3}^{\mathcal{R}}=D_{2}\left\{ \Omega _{2}^{\mathcal{R%
}}+\Omega _{3}^{\mathcal{R}}\right\} -\mathfrak{R}_{\mathcal{R}}(\mathfrak{T}%
_{2})+\mathfrak{R}_{\mathcal{R}}(\mathfrak{T}_{3}),  \label{2_} \\
(1-D_{1}-D_{2})\Omega _{3}^{\mathcal{R}}=D_{4}\left\{ \Omega _{4}^{\mathcal{R%
}}+\Omega _{3}^{\mathcal{R}}\right\} -\mathfrak{R}_{\mathcal{R}}(\mathfrak{T}%
_{4})+\mathfrak{R}_{\mathcal{R}}(\mathfrak{T}_{3}).  \label{3_}
\end{gather}%
Using Eqs. (\ref{1_}, \ref{2_}, \ref{3_}), we now express $D_{2}$ and $D_{4}$
in terms of $D_{1}$. For this, we subtract Eq (\ref{2_}) from Eq (\ref{1_})

\begin{equation}
(D_{1}-D_{2})\Omega _{3}^{\mathcal{R}}=D_{1}\left\{ \Omega _{1}^{\mathcal{R}%
}+\Omega _{3}^{\mathcal{R}}\right\} -D_{2}\left\{ \Omega _{2}^{\mathcal{R}%
}+\Omega _{3}^{\mathcal{R}}\right\} -\mathfrak{R}_{\mathcal{R}}(\mathfrak{T}%
_{1})+\mathfrak{R}_{\mathcal{R}}(\mathfrak{T}_{2}),  \notag
\end{equation}%
which gives

\begin{equation}
D_{2}=\frac{D_{1}\Omega _{1}^{\mathcal{R}}-\mathfrak{R}_{\mathcal{R}}(%
\mathfrak{T}_{1})+\mathfrak{R}_{\mathcal{R}}(\mathfrak{T}_{2})}{\Omega _{2}^{%
\mathcal{R}}}.  \label{D2_D1}
\end{equation}%
Similarly, subtracting Eq (\ref{3_}) from (\ref{1_}) results in

\begin{equation}
D_{4}=\frac{D_{1}\Omega _{1}^{\mathcal{R}}+\mathfrak{R}_{\mathcal{R}}(%
\mathfrak{T}_{4})-\mathfrak{R}_{\mathcal{R}}(\mathfrak{T}_{1})}{\Omega _{4}^{%
\mathcal{R}}}.  \label{D4_D1}
\end{equation}

Using Eqs. (\ref{D2_D1}, \ref{D4_D1}), the marginal $D_{3}$ can then be
expressed in terms of the marginal $D_{1}$ as

\begin{equation}
D_{3}=1-D_{1}[1+\Omega _{1}^{\mathcal{R}}(\frac{1}{\Omega _{2}^{\mathcal{R}}}%
+\frac{1}{\Omega _{4}^{\mathcal{R}}})]-\frac{\mathfrak{R}_{\mathcal{R}}(%
\mathfrak{T}_{2})-\mathfrak{R}_{\mathcal{R}}(\mathfrak{T}_{1})}{\Omega _{2}^{%
\mathcal{R}}}-\frac{\mathfrak{R}_{\mathcal{R}}(\mathfrak{T}_{4})-\mathfrak{R}%
_{\mathcal{R}}(\mathfrak{T}_{1})}{\Omega _{4}^{\mathcal{R}}}.  \label{D3_D1}
\end{equation}%
Eqs. (\ref{D2_D1}, \ref{D4_D1}, \ref{D3_D1}) represent the rational
behaviour of the $\mathcal{R}$ team, which the $\mathcal{B}$ team can now
exploit to optimize its defence strategy $\left\langle D_{\mathfrak{T}%
}\right\rangle $.

From Eqs. (\ref{Reward_functions}) the payoff function of the $\mathcal{B}$
team can be expressed as

\begin{equation}
\Pi _{\mathcal{B}}\{\left\langle D_{\mathfrak{T}}\right\rangle ,\left\langle 
{\normalsize A}_{\mathfrak{T}}\right\rangle \}=\sum\limits_{i=1}^{4}\left[
D_{i}\Omega _{i}^{\mathcal{B}}-\mathfrak{C}_{\mathcal{B}}(\mathfrak{T}_{i})%
\right] {\normalsize A}_{i},
\end{equation}%
and with the substitution $D_{3}=(1-D_{1}-D_{2}-D_{4})$ this result in
obtaining

\begin{gather}
\Pi _{\mathcal{B}}\{\left\langle D_{\mathfrak{T}}\right\rangle ,\left\langle 
{\normalsize A}_{\mathfrak{T}}\right\rangle \}=  \notag \\
=D_{1}[\Omega _{1}^{\mathcal{B}}{\normalsize A}_{1}-\Omega _{3}^{\mathcal{B}}%
{\normalsize A}_{3}]-\mathfrak{C}_{\mathcal{B}}(\mathfrak{T}_{1})%
{\normalsize A}_{1}+D_{2}[\Omega _{2}^{\mathcal{B}}{\normalsize A}%
_{2}-\Omega _{3}^{\mathcal{B}}{\normalsize A}_{3}]-\mathfrak{C}_{\mathcal{B}%
}(\mathfrak{T}_{2}){\normalsize A}_{2}  \notag \\
+D_{4}[\Omega _{4}^{\mathcal{B}}{\normalsize A}_{4}-\Omega _{3}^{\mathcal{B}}%
{\normalsize A}_{3}]-\mathfrak{C}_{\mathcal{B}}(\mathfrak{T}_{4})%
{\normalsize A}_{4}+\Omega _{3}^{\mathcal{B}}{\normalsize A}_{3}-\mathfrak{C}%
_{\mathcal{B}}(\mathfrak{T}_{3}){\normalsize A}_{3}.  \label{team_B_payoff}
\end{gather}%
Substitute from Eqs. (\ref{D2_D1}, \ref{D4_D1}) to Eq. (\ref{team_B_payoff}%
), along with the substitutions (\ref{substitutions}), to obtain

\begin{gather}
\Pi _{\mathcal{B}}\{\left\langle D_{\mathfrak{T}}\right\rangle ,\left\langle 
{\normalsize A}_{\mathfrak{T}}\right\rangle \}=  \notag \\
D_{1}\{\Omega _{1}^{\mathcal{B}}{\normalsize A}_{1}-\Omega _{3}^{\mathcal{B}}%
{\normalsize A}_{3}+\frac{\Omega _{1}^{\mathcal{R}}[\Omega _{2}^{\mathcal{B}}%
{\normalsize A}_{2}-\Omega _{3}^{\mathcal{B}}{\normalsize A}_{3}]}{\Omega
_{2}^{\mathcal{R}}}+\frac{\Omega _{1}^{\mathcal{R}}[\Omega _{4}^{\mathcal{B}}%
{\normalsize A}_{4}-\Omega _{3}^{\mathcal{B}}{\normalsize A}_{3}]}{\Omega
_{4}^{\mathcal{R}}}\}  \notag \\
+\frac{[-\mathfrak{R}_{\mathcal{R}}(\mathfrak{T}_{1})+\mathfrak{R}_{\mathcal{%
R}}(\mathfrak{T}_{2})][\Omega _{2}^{\mathcal{B}}{\normalsize A}_{2}-\Omega
_{3}^{\mathcal{B}}{\normalsize A}_{3}]}{\Omega _{2}^{\mathcal{R}}}  \notag \\
+\frac{[\mathfrak{R}_{\mathcal{R}}(\mathfrak{T}_{4})-\mathfrak{R}_{\mathcal{R%
}}(\mathfrak{T}_{1})][\Omega _{4}^{\mathcal{B}}{\normalsize A}_{4}-\Omega
_{3}^{\mathcal{B}}{\normalsize A}_{3}]}{\Omega _{4}^{\mathcal{R}}}  \notag \\
+\Omega _{3}^{\mathcal{B}}{\normalsize A}_{3}-[\mathfrak{C}_{\mathcal{B}}(%
\mathfrak{T}_{1}){\normalsize A}_{1}+\mathfrak{C}_{\mathcal{B}}(\mathfrak{T}%
_{2}){\normalsize A}_{2}+\mathfrak{C}_{\mathcal{B}}(\mathfrak{T}_{3})%
{\normalsize A}_{3}+\mathfrak{C}_{\mathcal{B}}(\mathfrak{T}_{4}){\normalsize %
A}_{4}]  \notag \\
=D_{1}\Delta _{1}+\Delta _{2}  \label{B_team_payoffs}
\end{gather}%
where

\begin{eqnarray}
\Delta _{1} &=&\Omega _{1}^{\mathcal{B}}{\normalsize A}_{1}-\Omega _{3}^{%
\mathcal{B}}{\normalsize A}_{3}+\frac{\Omega _{1}^{\mathcal{R}}[\Omega _{2}^{%
\mathcal{B}}{\normalsize A}_{2}-\Omega _{3}^{\mathcal{B}}{\normalsize A}_{3}]%
}{\Omega _{2}^{\mathcal{R}}}+\frac{\Omega _{1}^{\mathcal{R}}[\Omega _{4}^{%
\mathcal{B}}{\normalsize A}_{4}-\Omega _{3}^{\mathcal{B}}{\normalsize A}_{3}]%
}{\Omega _{4}^{\mathcal{R}}}  \notag \\
&&  \label{delta_1} \\
\Delta _{2} &=&\frac{[-\mathfrak{R}_{\mathcal{R}}(\mathfrak{T}_{1})+%
\mathfrak{R}_{\mathcal{R}}(\mathfrak{T}_{2})][\Omega _{2}^{\mathcal{B}}%
{\normalsize A}_{2}-\Omega _{3}^{\mathcal{B}}{\normalsize A}_{3}]}{\Omega
_{2}^{\mathcal{R}}}+\frac{[\mathfrak{R}_{\mathcal{R}}(\mathfrak{T}_{4})-%
\mathfrak{R}_{\mathcal{R}}(\mathfrak{T}_{1})][\Omega _{4}^{\mathcal{B}}%
{\normalsize A}_{4}-\Omega _{3}^{\mathcal{B}}{\normalsize A}_{3}]}{\Omega
_{4}^{\mathcal{R}}}  \notag \\
&&+\Omega _{3}^{\mathcal{B}}{\normalsize A}_{3}-[\mathfrak{C}_{\mathcal{B}}(%
\mathfrak{T}_{1}){\normalsize A}_{1}+\mathfrak{C}_{\mathcal{B}}(\mathfrak{T}%
_{2}){\normalsize A}_{2}+\mathfrak{C}_{\mathcal{B}}(\mathfrak{T}_{3})%
{\normalsize A}_{3}+\mathfrak{C}_{\mathcal{B}}(\mathfrak{T}_{4}){\normalsize %
A}_{4}],  \label{delta_2}
\end{eqnarray}%
appear as the new parameters of considered sequential strategic interaction.
This completes the backwards induction process of obtaining the optimal
response of the $\mathcal{B}$ team in view of its encounter with the
rational behaviour of the $\mathcal{R}$ team.

\section{Optimal response of the $\mathcal{B}$ team}

From Eq. (\ref{delta_1}, \ref{delta_2}) we note that $\Delta _{1,2}$ depend
on the values assigned to the two teams' rewards and costs variables i.e. $%
\mathfrak{R}_{\mathcal{B}}(\mathfrak{T}),$ $\mathfrak{C}_{\mathcal{B}}(%
\mathfrak{T}),$ $\mathfrak{R}_{\mathcal{R}}(\mathfrak{T}),$ $\mathfrak{C}_{%
\mathcal{R}}(\mathfrak{T})$ as well as on the $\mathcal{R}$ team's attack
probabilities ${\normalsize A}_{i}(1\leq i\leq 4).$ Three case, therefore,
emerge in view of Eq. (\ref{B_team_payoffs}) that are described below.

\subsection{Case $\Delta _{1}>0$}

After observing the attack probabilities ${\normalsize A}_{i}(1\leq i\leq 4)$
the $\mathcal{B}$ team obtains $\Delta _{1}$ using Eq. (\ref{delta_1}) along
with the rewards and costs variables i.e. $\mathfrak{R}_{\mathcal{B}}(%
\mathfrak{T}),$ $\mathfrak{C}_{\mathcal{B}}(\mathfrak{T}),$ $\mathfrak{R}_{%
\mathcal{R}}(\mathfrak{T}),$ $\mathfrak{C}_{\mathcal{R}}(\mathfrak{T})$ and
the Eqs. (\ref{substitutions}). If the $\mathcal{B}$ team finds that $\Delta
_{1}>0$ then its payoff $\Pi _{\mathcal{B}}\{\left\langle D_{\mathfrak{T}%
}\right\rangle ,\left\langle {\normalsize A}_{\mathfrak{T}}\right\rangle \}$
is maximized at the maximum value of $D_{1}$ and it is irrespective of the
value of $\Delta _{2}$. Note that at this maximum value of $D_{1}$, the
corresponding values of $D_{2},D_{3},D_{4}$--- as expressed in terms of $%
D_{1}$ and given by Eqs. (\ref{D2_D1}, \ref{D3_D1}, \ref{D4_D1})--- must
remain non-negative, and that the maximum value obtained for $D_{1}$ can
still be less than $D_{2}$ or $D_{3}$ or $D_{4}.$

\subsection{Case $\Delta _{1}<0$}

As $0\leq D_{i}\leq 1$ and $\sum\limits_{i=1}^{4}D_{i}=1$, therefore, in
view of the attack probabilities ${\normalsize A}_{i}(1\leq i\leq 4)$, if
the $\mathcal{B}$ team finds that $\Delta _{1}<0$ then the reward is
maximized to the value of $\Delta _{2}$ with $D_{1}=0$ and we then have

\begin{eqnarray}
D_{2} &=&\frac{\mathfrak{R}_{\mathcal{R}}(\mathfrak{T}_{2})-\mathfrak{R}_{%
\mathcal{R}}(\mathfrak{T}_{1})}{\Omega _{2}^{\mathcal{R}}},  \notag \\
\text{ }D_{3} &=&1-\frac{\mathfrak{R}_{\mathcal{R}}(\mathfrak{T}_{2})-%
\mathfrak{R}_{\mathcal{R}}(\mathfrak{T}_{1})}{\Omega _{2}^{\mathcal{R}}}-%
\frac{\mathfrak{R}_{\mathcal{R}}(\mathfrak{T}_{4})-\mathfrak{R}_{\mathcal{R}%
}(\mathfrak{T}_{1})}{\Omega _{4}^{\mathcal{R}}},  \notag \\
\text{ }D_{4} &=&\frac{\mathfrak{R}_{\mathcal{R}}(\mathfrak{T}_{4})-%
\mathfrak{R}_{\mathcal{R}}(\mathfrak{T}_{1})}{\Omega _{4}^{\mathcal{R}}}.
\label{Ds_when_D1_is_zero}
\end{eqnarray}

\subsection{Case $\Delta _{1}=0$}

If the $\mathcal{B}$ team finds that $\Delta _{1}=0$ then its reward becomes 
$\Delta _{2}$, as defined by Eq. (\ref{delta_2}), and is independent of the
value assigned to $D_{1}$ and via Eqs. (\ref{D2_D1}, \ref{D3_D1}, \ref{D4_D1}%
) also independent of $D_{2},$ $D_{3},$ $D_{4}$.

\section{Example instantiation}

As an example, we consider the set of arbitrarily-assigned values to the two
teams' rewards and costs as in the table below.

\begin{equation}
\begin{array}{c}
\text{Tanks}%
\end{array}%
\begin{tabular}{|l|}
\hline
$T$ \\ \hline
$\mathfrak{T}_{1}$ \\ \hline
$\mathfrak{T}_{2}$ \\ \hline
$\mathfrak{T}_{3}$ \\ \hline
$\mathfrak{T}_{4}$ \\ \hline
\end{tabular}%
\overset{%
\begin{array}{c}
\text{An example of rewards}%
\end{array}%
}{\overset{%
\begin{array}{c}
\mathcal{B}\text{ team}%
\end{array}%
}{%
\begin{tabular}{|l|l|}
\hline
$\mathfrak{R}_{\mathcal{B}}$ & $\mathfrak{C}_{\mathcal{B}}$ \\ \hline
$9$ & $7$ \\ \hline
$8$ & $7$ \\ \hline
$7$ & $6$ \\ \hline
$4$ & $2$ \\ \hline
\end{tabular}%
}%
\begin{array}{c}
\\ 
\\ 
\\ 
\\ 
\end{array}%
\begin{array}{c}
\\ 
\\ 
\\ 
\\ 
\end{array}%
\begin{array}{c}
\\ 
\\ 
\\ 
\\ 
\end{array}%
\overset{%
\begin{array}{c}
\mathcal{R}\text{ team}%
\end{array}%
}{%
\begin{tabular}{|l|l|}
\hline
$\mathfrak{R}_{\mathcal{R}}$ & $\mathfrak{C}_{\mathcal{R}}$ \\ \hline
$4$ & $2$ \\ \hline
$6$ & $5$ \\ \hline
$7$ & $3$ \\ \hline
$5$ & $1$ \\ \hline
\end{tabular}%
}},  \label{table}
\end{equation}%
for which we have

\begin{eqnarray}
\Omega _{1}^{\mathcal{B}} &=&16,\text{ }\Omega _{1}^{\mathcal{R}}=6,\text{ }%
\Omega _{2}^{\mathcal{B}}=15,\text{ }\Omega _{2}^{\mathcal{R}}=11,  \notag \\
\Omega _{3}^{\mathcal{B}} &=&13,\text{ }\Omega _{3}^{\mathcal{R}}=10,\text{ }%
\Omega _{4}^{\mathcal{B}}=6,\text{ }\Omega _{4}^{\mathcal{R}}=6,
\end{eqnarray}%
and for which using Eqs. (\ref{delta_1}, \ref{delta_2}) we obtain

\begin{eqnarray}
\Delta _{1} &=&16{\normalsize A}_{1}+8.181{\normalsize A}_{2}-33.090%
{\normalsize A}_{3}+6{\normalsize A}_{4},  \label{delta_1_example} \\
\Delta _{2} &=&-\mathfrak{7}{\normalsize A}_{1}-4.272{\normalsize A}%
_{2}+2.469{\normalsize A}_{3}-{\normalsize A}_{4}.  \label{delta_2_example}
\end{eqnarray}

\subsection{Case $\Delta _{1}>0$}

Now, assume that while knowing the attack probabilities ${\normalsize A}%
_{i}(1\leq i\leq 4)$, the $\mathcal{B}$ team uses Eq. (\ref{delta_1_example}%
) to find that $\Delta _{1}>0.$ As discussed above, its payoff is maximized
at the maximum value of $D_{1}$ and it is irrespective of the value of $%
\Delta _{2}.$ Using Eqs. (\ref{D2_D1}, \ref{D4_D1}), along with the enteries
in the table (\ref{table}), the $\mathcal{B}$ team now determines the
maximum value for $D_{1}$ at which $D_{2},$ $D_{3},$ $D_{4}$ obtained from
Eqs. (\ref{D2_D1}, \ref{D3_D1}, \ref{D4_D1}), respectively, all have
non-negative values. Table (\ref{table}) gives

\begin{equation}
D_{2}=\frac{6D_{1}+2}{11},\text{ }D_{3}=1-10.272D_{1}-0.3485,\text{ }D_{4}=%
\frac{6D_{1}+\mathfrak{1}}{6},  \label{Ds_from_D1}
\end{equation}%
and a table of values is then obtained as

\begin{equation}
\begin{tabular}{|l|l|l|l|}
\hline
$D_{1}$ & $D_{2}$ & $D_{3}$ & $D_{4}$ \\ \hline
$0.0632$ & $0.216$ & $0.002$ & $0.23$ \\ \hline
$0.0633$ & $0.216$ & $0.001$ & $0.23$ \\ \hline
$0.0634$ & $0.216$ & $2.24\cdot 10^{-4}$ & $0.23$ \\ \hline
$0.0635$ & $0.216$ & $-8.03\cdot 10^{-4}$ & $0.23$ \\ \hline
$0.0636$ & $0.217$ & $-0.002$ & $0.23$ \\ \hline
\end{tabular}%
,
\end{equation}%
and $D_{1}=0.0634$ emerges as the maximum value at which $D_{2},$ $D_{3},$ $%
D_{4}$ remain non-negative.

\begin{figure}[tbp]
\centering
\includegraphics[width=1.0\linewidth]{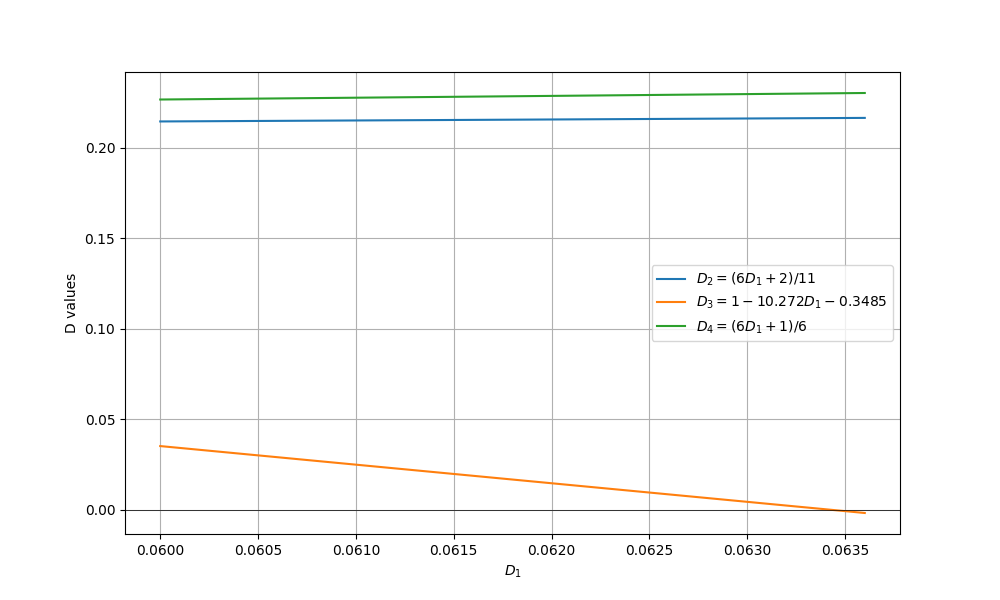}
\caption{Plots of $D_2$, $D_3$, and $D_4$ vs $D_1$ (Range: 0.06 to 0.0636)}
\label{fig:enter-label}
\end{figure}

The $\mathcal{B}$ team's protection strategy is therefore obtained from Eqs.
(\ref{Ds_from_D1}) as

\begin{equation}
\left\langle D_{\mathfrak{T}}\right\rangle ^{\ast }=\left\langle D_{1}^{\ast
},\text{ }D_{2}^{\ast },\text{ }D_{3}^{\ast },\text{ }D_{4}^{\ast
}\right\rangle =\left\langle 0.0634,\text{ }0.216,\text{ }2.24\cdot 10^{-4},%
\text{ }0.23\right\rangle ,  \label{Protection_Strategy}
\end{equation}%
and from Eq. (\ref{B_team_payoffs}, \ref{delta_1_example}, \ref%
{delta_2_example}), along with the table (\ref{table}), the $\mathcal{B}$
team's payoffs then become

\begin{equation}
\Pi _{\mathcal{B}}\{\left\langle D_{\mathfrak{T}}\right\rangle ^{\ast
},\left\langle {\normalsize A}_{\mathfrak{T}}\right\rangle \}=-5.986%
{\normalsize A}_{1}-3.753{\normalsize A}_{2}+0.371A_{3}-0.62A_{4}
\label{B_team_payoff_1}
\end{equation}%
which, in view of the fact that $\sum\limits_{i=1}^{4} A_{i} = 1$, can also be expressed as

\begin{equation}
\Pi _{\mathcal{B}}\{\left\langle D_{\mathfrak{T}}\right\rangle ^{\ast
},\left\langle {\normalsize A}_{\mathfrak{T}}\right\rangle
\}=-6.357A_{1}-4.124A_{2}-0.991A_{4}+0.371.  \label{B_team_payoff_2}
\end{equation}

Plot of the payoff $\Pi _{\mathcal{B}%
}\{\left\langle D_{\mathfrak{T}}\right\rangle ^{\ast },\left\langle 
{\normalsize A}_{\mathfrak{T}}\right\rangle \}$ for the values of $A_{1},$ $%
A_{2},$ $A_{4}$ that satisfy the constraints $0\leq A_{1},$ $A_{2},$ $%
A_{4}\leq 1$ and $0\leq {\normalsize A}_{1}+{\normalsize A}_{2}+{\normalsize %
A}_{4}\leq 1$ is in FIG. 2.

\begin{figure}[tbp]
\centering
\includegraphics[width=1.0\linewidth]{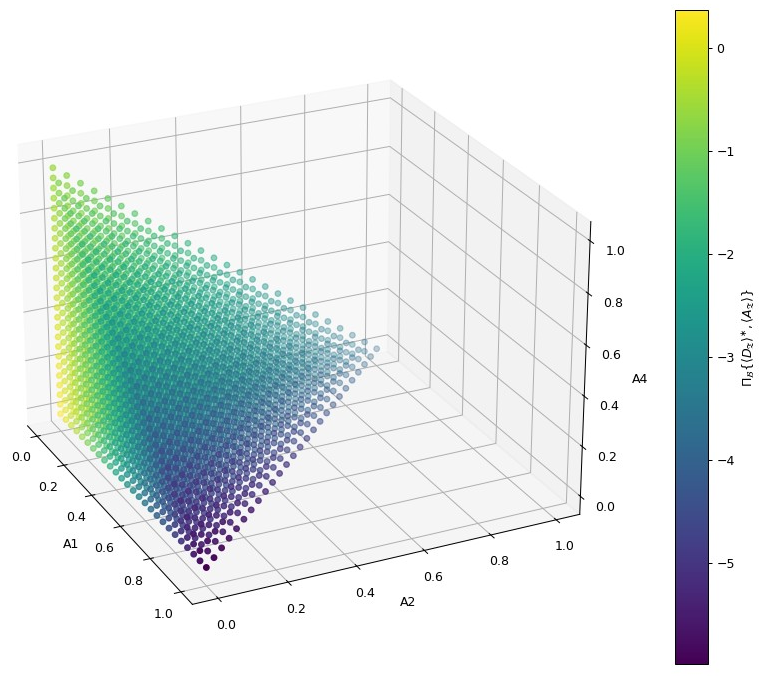}
\caption{The $\mathcal{B}$ team's payoff as given by Eq. (\protect\ref%
{B_team_payoff_2}) when $\Delta _{1}>0$ and for the values of $A_{1},$ $%
A_{2},$ $A_{4}$ that satisfy the constraints $0\leq A_{1},$ $A_{2},$ $%
A_{4}\leq 1$ and $0\leq {\protect\normalsize A}_{1}+{\protect\normalsize A}%
_{2}+{\protect\normalsize A}_{4}\leq 1.$}
\label{fig:enter-label}
\end{figure}

From Eq. (\ref{B_team_payoff_2}) the payoff $\Pi _{\mathcal{B}%
}\{\left\langle D_{\mathfrak{T}}\right\rangle ^{\ast },\left\langle 
{\normalsize A}_{\mathfrak{T}}\right\rangle \}$ is maximized at the value of 
$0.371$ for $A_{3}=1,$ and therefore $A_{1,2,4}=0.$ The payoff to the $%
\mathcal{R}$ team is then obtained from Eq. (\ref{R_team_payoff}) as

\begin{equation}
\Pi _{\mathcal{R}}\{\left\langle D_{\mathfrak{T}}\right\rangle ^{\ast
},\left\langle {\normalsize A}_{\mathfrak{T}}\right\rangle \}=(1-D_{3}^{\ast
})\mathfrak{R}_{\mathcal{R}}(\mathfrak{T}_{3})-D_{3}^{\ast }\mathfrak{C}_{%
\mathcal{R}}(\mathfrak{T}_{3}),
\end{equation}%
where from Eq. (\ref{Protection_Strategy}) we have $D_{3}^{\ast }\approx 0$
and using the table (\ref{table}) we obtain $\Pi _{\mathcal{R}%
}\{\left\langle D_{\mathfrak{T}}\right\rangle ^{\ast },\left\langle 
{\normalsize A}_{\mathfrak{T}}\right\rangle \}=\mathfrak{R}_{\mathcal{R}}(%
\mathfrak{T}_{3})=7.$

Now we consider the reaction of the $\mathcal{R}$ team after the $\mathcal{B}
$ team has determined its protection strategy $\left\langle D_{\mathfrak{T}%
}\right\rangle ^{\ast }$ while following the backwards induction in the case
above. For the case when the attack probabilities are such that $\Delta
_{1}>0$ in Eq. (\ref{delta_1_example}), we reexpress the $\mathcal{R}$
team's payoff given by Eq. (\ref{R_team_payoff}) by substituting the $%
\mathcal{B}$ team's protection strategy described by Eq. (\ref%
{Protection_Strategy}). Using $\sum\limits_{i=1}^{4}{\normalsize A}_{i}=1$,
the $\mathcal{R}$ team's payoff are then expressed in terms of the attack
probabilities $A_{1},A_{2},A_{4}$ as

\begin{equation}
\Pi _{\mathcal{R}}\{\left\langle D_{\mathfrak{T}}\right\rangle ^{\ast
},\left\langle {\normalsize A}_{\mathfrak{T}}\right\rangle \}=-3.38(%
{\normalsize A}_{1}+{\normalsize A}_{2}+{\normalsize A}_{4})+\mathfrak{7}.
\label{R_team_payoff_1}
\end{equation}%
Now, in FIG. 3 below a plot is obtained comparing the $\mathcal{B}$ and the $%
\mathcal{R}$ team's payoffs given by Eqs. (\ref{B_team_payoff_2}, \ref%
{R_team_payoff_1}), respectively, when these are considered implicit
functions of $A_{1},$ $A_{2},$ $A_{4}$ and with the constraints that $0\leq
A_{1},$ $A_{2},$ $A_{4}\leq 1$ and $0\leq {\normalsize A}_{1}+{\normalsize A}%
_{2}+{\normalsize A}_{4}\leq 1.$ For most of the allowed values of the
attack probabilities, as represented by the blue shade, and for $\Delta
_{1}>0$, $%
\mathcal{R}$ team remains significantly better off than the $\mathcal{B}$ team. 

\begin{figure}[tbp]
\centering
\includegraphics[width=1\linewidth]{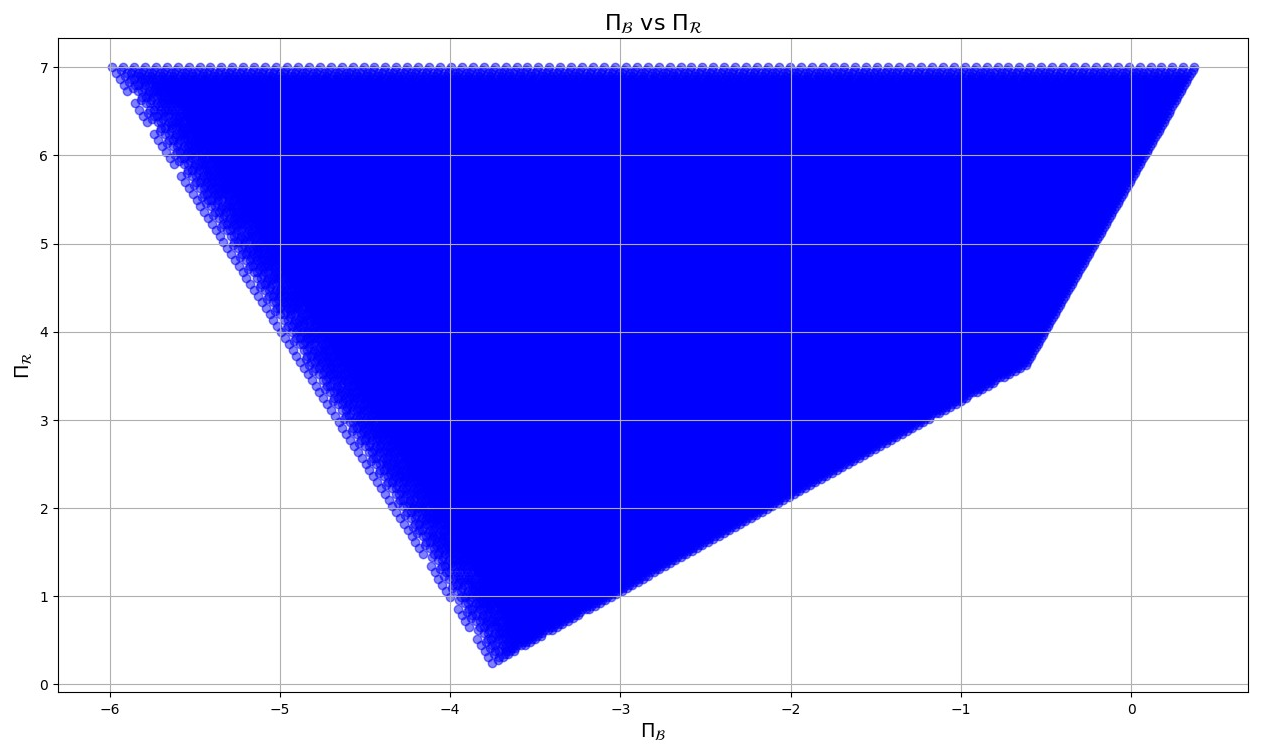}
\caption{The plot between $\Pi _{\mathcal{B}}$ and $\Pi _{\mathcal{R}}$ when 
$\Delta _{1}>0$ and for values of $A_{1},$ $A_{2},$ $A_{4}$ that satisfy the
constraints $0\leq A_{1},$ $A_{2},$ $A_{4}\leq 1$ and $0\leq 
{\protect\normalsize A}_{1}+{\protect\normalsize A}_{2}+{\protect\normalsize %
A}_{4}\leq 1.$}
\label{fig:enter-label}
\end{figure}

In view of the reward table (\ref{table}), the $\mathcal{R}$ team's payoffs
attain the maximum value of $7$ when $A_{1}+A_{2}+A_{4}=0$ or when $A_{3}=1.$
However, when this is the case, using Eq. (\ref{B_team_payoff_2}) the payoff
to the $\mathcal{B}$ team then becomes $0.371$.

\subsection{Case $\Delta _{1}\leq 0$}

Consider the case when by using Eq. (\ref{delta_1_example}) the $\mathcal{B}$
team finds that $\Delta _{1}\leq 0$. As $0\leq {\normalsize A}_{1}+%
{\normalsize A}_{2}+{\normalsize A}_{4}\leq 1$ in Eq. (\ref{delta_1_example}%
), the condition $\Delta _{1}\leq 0$ can be realized for some values of the
attack probabilities.

Plot of the payoff $\Pi _{\mathcal{B}%
}\{\left\langle D_{\mathfrak{T}}\right\rangle ^{\ast },\left\langle 
{\normalsize A}_{\mathfrak{T}}\right\rangle \}$ for the values of $A_{1},$ $%
A_{2},$ $A_{4}$ that satisfy the constraints $0\leq A_{1},$ $A_{2},$ $%
A_{4}\leq 1$ and $0\leq {\normalsize A}_{1}+{\normalsize A}_{2}+{\normalsize %
A}_{4}\leq 1$ is in FIG. 4.

Now in view of Eq. (\ref{B_team_payoffs}) the $%
\mathcal{B}$ team's reward is maximized to the value of $\Delta _{2}$ when $%
D_{1}=0.$ In this case, using Eq. (\ref{Ds_when_D1_is_zero}) and the table (%
\ref{table}) the $\mathcal{B}$ team's protection strategy is therefore
obtained as

\begin{equation}
\left\langle D_{\mathfrak{T}}\right\rangle ^{\ast }=\left\langle D_{1}^{\ast
},\text{ }D_{2}^{\ast },\text{ }D_{3}^{\ast },\text{ }D_{4}^{\ast
}\right\rangle =\left\langle 0,\text{ }0.181,\text{ }0.651,\text{ }%
0.167\right\rangle ,  \label{Protection_Strategy_b}
\end{equation}%
and, as before, using Eq. (\ref{B_team_payoffs}, \ref{delta_1_example}, \ref%
{delta_2_example}), along with the table (\ref{table}), the $\mathcal{B}$
team's payoffs then become $\Pi _{\mathcal{B}}\{\left\langle D_{\mathfrak{T}%
}\right\rangle ^{\ast },\left\langle {\normalsize A}_{\mathfrak{T}%
}\right\rangle \}=\Delta _{2}$ i.e.

\begin{equation}
\Pi _{\mathcal{B}}\{\left\langle D_{\mathfrak{T}}\right\rangle ^{\ast
},\left\langle {\normalsize A}_{\mathfrak{T}}\right\rangle \}=-7A_{1}-4.272%
{\normalsize A}_{2}+2.47A_{3}-A_{4},  \label{B_team_payoff_at_star}
\end{equation}%
which in view of the fact that $\displaystyle\sum_{i=1}^{4} A_{i}=1$
%\sum\limits_{i=1}^{4}{\normalsize A}_{i}=1
can also be expressed as

\begin{equation}
\Pi _{\mathcal{B}}\{\left\langle D_{\mathfrak{T}}\right\rangle ^{\ast
},\left\langle {\normalsize A}_{\mathfrak{T}}\right\rangle \}=-9.47%
{\normalsize A}_{1}-6.742{\normalsize A}_{2}-3.47{\normalsize A}_{4}+2.47,
\end{equation}%
and similarly for the $\mathcal{R}$ team

\begin{equation}
\Pi _{\mathcal{R}}\{\left\langle D_{\mathfrak{T}}\right\rangle ^{\ast
},\left\langle {\normalsize A}_{\mathfrak{T}}\right\rangle \}=3.51(%
{\normalsize A}_{1}+{\normalsize A}_{2}+{\normalsize A}_{4})+0.49.
\end{equation}

\begin{figure}[tbp]
\centering
\includegraphics[width=1\linewidth]{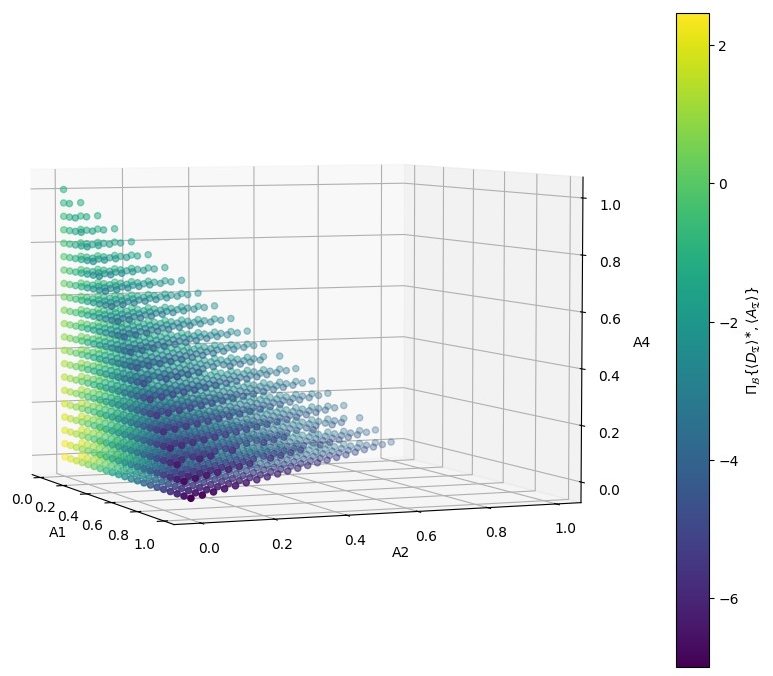}
\caption{The $\mathcal{B}$ team's payoff as given by Eq. (\protect\ref%
{B_team_payoff_2}) when $\Delta _{1}\leq 0$ and for the values of $A_{1},$ $%
A_{2},$ $A_{4}$ that satisfy the constraints $0\leq A_{1},$ $A_{2},$ $%
A_{4}\leq 1$ and $0\leq {\protect\normalsize A}_{1}+{\protect\normalsize A}%
_{2}+{\protect\normalsize A}_{4}\leq 1.$}
\label{fig:enter-label}
\end{figure}

\begin{figure}[tbp]
\centering
\includegraphics[width=1\linewidth]{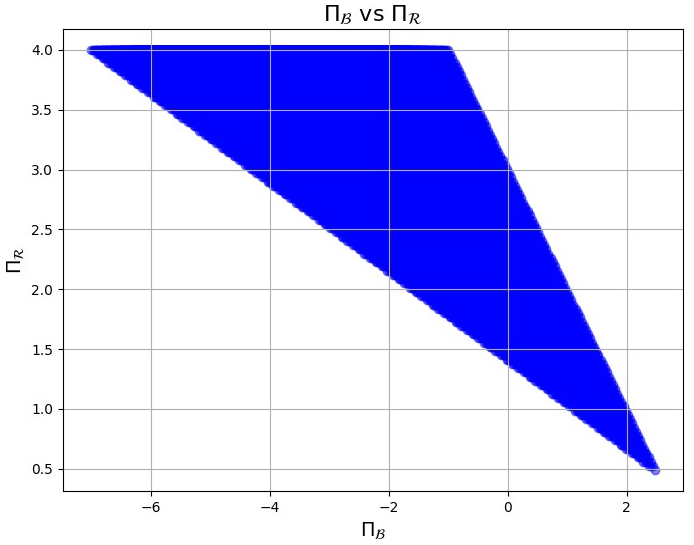}
\caption{The plot between $\Pi _{\mathcal{B}}$ and $\Pi _{\mathcal{R}}$ when 
$\Delta _{1}\leq 0$ and for values of $A_{1},$ $A_{2},$ $A_{4}$ that satisfy
the constraints $0\leq A_{1},$ $A_{2},$ $A_{4}\leq 1$ and $0\leq 
{\protect\normalsize A}_{1}+{\protect\normalsize A}_{2}+{\protect\normalsize %
A}_{4}\leq 1.$}
\label{fig:enter-label}
\end{figure}

\section{Discussion}

We consider the case that the $\mathcal{B}$ team moves first and commits to
a protection strategy $\left\langle D_{\mathfrak{T}}\right\rangle $. The $%
\mathcal{R}$ team notices the protection strategy $\left\langle D_{\mathfrak{%
T}}\right\rangle $ first and then decides its attack strategy given by the
vector $\left\langle {\normalsize A}_{\mathfrak{T}}\right\rangle $. The $%
\mathcal{B}$ team knows that the $\mathcal{R}$ team is observing its actions
defining the protection strategy $\left\langle D_{\mathfrak{T}}\right\rangle 
$. The $\mathcal{B}$ team also knows that the $\mathcal{R}$ team is a
rational decision maker and how it will react to its protection strategy $%
\left\langle D_{\mathfrak{T}}\right\rangle $. The leader-follower
interaction---resulting in the consideration of Stackelberg
equilibrium---looks into finding the $\mathcal{B}$ team's best protection
strategy $\left\langle D_{\mathfrak{T}}\right\rangle ^{\ast }$ while knowing
that the $\mathcal{R}$ team is going to act rationally in view of a
protection strategy $\left\langle D_{\mathfrak{T}}\right\rangle $ committed
by the $\mathcal{B}$ team. The $\mathcal{R}$ team's mixed strategy is given
by the vector $\left\langle {\normalsize A}_{\mathfrak{T}}\right\rangle $ of
the the attack probabilities $A_{1},A_{2},A_{3},A_{4}$ on the four tanks by
the $\mathcal{R}$ team.

The vector $\left\langle D_{\mathfrak{T}}\right\rangle $ describing the $%
\mathcal{B}$ team's allocation of its resources depends crucially on the
parameter $\Delta _{1}$ as defined in Eq. (\ref{3_}) and is obtained from
the assigned values in table (\ref{table}) for rewards and the costs to the
two teams. If $\Delta _{1}>0$ then the reward to the $\mathcal{B}$ team is
maximized at the maximum value of $D_{1}$ for which $D_{2},D_{3},D_{4}$---
as expressed in terms of $D_{1}$ by Eqs. (\ref{D2_D1}, \ref{D3_D1}, \ref%
{D4_D1})---remain non-negative. That is, the maximum value obtained for $%
D_{1}$ can still be less than $D_{2}$ or $D_{3}$ or $D_{4}.$

We note that for the case $\Delta _{1}>0$ and for most situations
encountered by the two teams---represented by the area covered by the blue
shade in FIG. 3---the reward for the $\mathcal{B}$ team remains between $-6$
and $0.5$ whereas the reward for the $\mathcal{R}$ team remains between $0.5$
and $7$.

However, in the case $\Delta _{1}\leq 0$ and for most situations encountered
by the two teams---represented now by the area covered by the blue shade in
FIG. 5---the reward for the $\mathcal{B}$ team remains between $-6.5$ and $%
2.5$ whereas the reward for the $\mathcal{R}$ team remains between $0.5$ and 
$4$.

That is, for most of the allowed values of the attack probabilities, the $%
\mathcal{B}$ team can receive higher reward when $\Delta _{1}\leq 0$
relative to the case when $\Delta _{1}>0.$ However, for most of the allowed
values of the attack probabilities, the $\mathcal{R}$ team can receive less
reward when $\Delta _{1}\leq 0$ relative to the case when $\Delta _{1}>0.$
Therefore, the situation $\Delta _{1}\leq 0$ is more favorable to the $%
\mathcal{B}$ team than it is to the $\mathcal{R}$ team. Similarly, the
situation $\Delta _{1}>0$ turns out to be more favorable to the $\mathcal{R}$
team than it is to the $\mathcal{B}$ team. Note that these results are
specific to the particular values assigned in the considered example to the
parameters $\mathfrak{R}_{\mathcal{B}}(\mathfrak{T}),$ $\mathfrak{C}_{%
\mathcal{B}}(\mathfrak{T}),$ $\mathfrak{R}_{\mathcal{R}}(\mathfrak{T}),$ $%
\mathfrak{C}_{\mathcal{R}}(\mathfrak{T})$ and for four tanks only.

\section{Conclusions}

The Stackelberg equilibrium in this scenario is reached when the drones have
optimized their surveillance and attack patterns, considering the tanks'
best responses, and the tanks have subsequently optimized their defensive
and offensive strategies in light of the drone tactics. This equilibrium
reached in the setting of sequential decision making represents the point
where neither side can unilaterally improve its position given the strategy
of the other. In this high-tech game of cat and mouse, the effectiveness of
each side hinges not only on their technological capabilities but also on
their ability to out think and outmaneuver the opponent within the
game-theoretic framework. The dynamic interplay of strategic
decision-making, under the principles of Stackelberg equilibrium and
backwards induction, highlights the intricate nature of modern warfare
involving drones and tanks where brains and brawn are equally pivotal. A
natural extension of this work is the case when the set of resources
consists of $m$ element i.e. $S=\{\mathcal{S}_{1},\mathcal{S}_{2},...%
\mathcal{S}_{m}\}$ and set of tanks is given as $T=\{\mathfrak{T}_{1},%
\mathfrak{T}_{2},...\mathfrak{T}_{n}\}$.

\section{Acknowledgement}
The work in this paper was carried out under a Research Agreement between the Defence
Science and Technology Group, Department of Defence, Australia, and the University of
Adelaide, Contract No. UA216424-S27.

\end{document}